\documentclass[aps,prl,preprint,amsmath,amssymb,groupedaddress]{revtex4-1}\usepackage{graphicx}% Include figure files
\usepackage{epstopdf}
\usepackage{dcolumn}% Align table columns on decimal point
\usepackage{bm}% bold math
\begin{document}
\preprint{Published in PRL, 111, 053201 (2013)}
\title{Quantum tunneling of oxygen atoms on very cold surfaces}% Force line breaks with \\
   \author{M.~Minissale}
\email{email adress: marco.minissale@obspm.fr}
\author{E. Congiu}%,
\author{S. Baouche}%
\author{H. Chaabouni}%
\author{A. Moudens}%
\author{F. Dulieu}%
\affiliation{%
Universit\'e de Cergy Pontoise and Observatoire de Paris, ENS, UPMC,
UMR 8112 du CNRS\\ 5, mail Gay Lussac, 95000 Cergy Pontoise cedex,
France.
}%
\author{M. Accolla}
\affiliation{Dipartimento di Scienze Applicate, Universit\`a degli Studi di Napoli
Parthenope, Centro Direzionale Isola C4, 80143 Napoli, Italy.
}%
\author{S. Cazaux}
\affiliation{Kapteyn Astronomical Institute, PO box 800, 9700AV Groningen, The Netherlands.
}%
\author{G. Manic\'o}
\author{V. Pirronello}
\affiliation{ Dipartimento di Fisica e Astronomia, Universit\`a di Catania, via Santa Sofia 64, 95123 Catania, Sicily, Italy.
}%
\begin{abstract}
%---213 words
Any evolving system can change of state via thermal mechanisms
(hopping a barrier) or via quantum tunneling. Most of the time,
efficient classical mechanisms dominate at high temperatures. This
is why an increase of the temperature can initiate the chemistry. We
present here an experimental investigation of O-atom diffusion and
reactivity on water ice. We explore the 6-25~K temperature range at
sub-monolayer surface coverages. We derive the diffusion temperature
law and observe the transition from quantum to classical diffusion. Despite of the high
mass of O, quantum tunneling is efficient even at 6~K. As a
consequence, the solid-state astrochemistry of cold regions should
be reconsidered and should include the possibility of forming larger
organic molecules than previously expected.
%\begin{description}
%\end{description}
\end{abstract}
\maketitle
%\tableofcontents
\section{Introduction}
\begin{figure}[t]
\centering
\includegraphics[width=7.5cm]{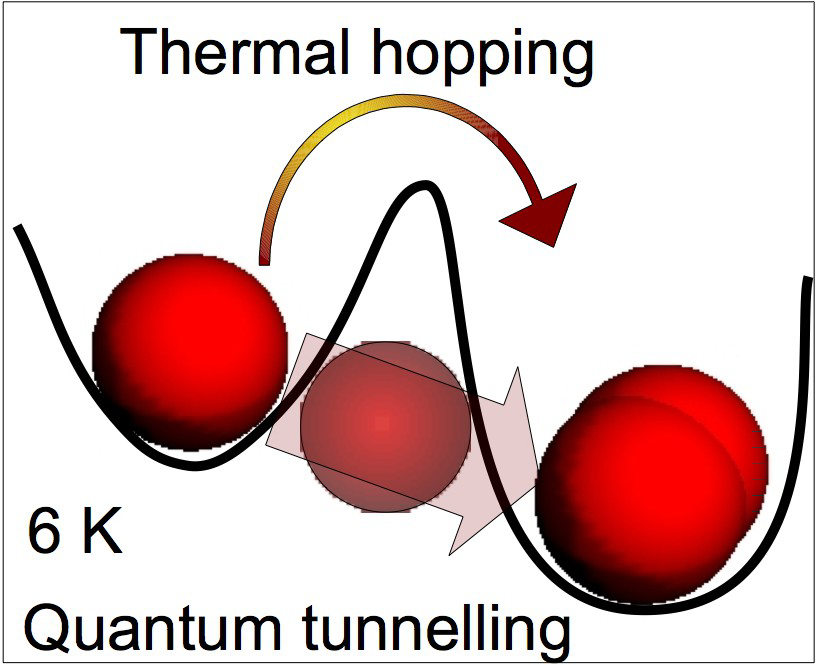}
\caption{Cartoon of thermal motion of an oxygen atom, vs quantum tunneling. Thermally-induced
diffusion follows the regular exponential Arrhenius Law with T, quantum diffusion does
not.} \label{fig:Tunn}
\end{figure}

\nocite{*} Nuclear decay or chemical reactions may be described as
the crossing through a potential barrier by quantum tunneling, or
crossing over the same barrier by thermal hopping (\figurename~\ref{fig:Tunn}). Except for barrier-less reactions,
increasing the temperature initiates the chemistry. Actually, the
quantum tunneling regime (like in nuclear decay), and the thermal
activation (like in chemistry) are usually separated by orders of
magnitude in temperature. Theoretically, the balance between
classical thermal motion and quantum tunneling is a very active
subject, especially because it impacts solid state chemistry
\cite{Go10}. It is well established that a critical temperature
exists below which tunneling is dominant (\cite{Me73},\cite{Gi87}).
Experimentally, such an evidence is still missing in handbooks.
Thanks to field ion microscopy, diffusion of single atoms on metals
or on crystalline surfaces has been studied in detail for decades
\cite{Eh94}. However, on amorphous surfaces, and especially on
water ice substrates, the study of physisorbed light atoms presents
enormous difficulties for any atomic microscopy technique that the
field is still nearly unexplored. Yet, in the cold regions of the
Universe, where temperatures are lower than 8K \cite{Pa07}, a
rich chemistry is initiated on the surfaces of minuscule dust
particles \cite{Ti82}. The species weakly bound to the surface are
the pivotal media of this pristine chemistry, governed by the
diffusion of reactive species \cite{Cu07}. So far, diffusion has
only been partly explored experimentally for H atoms
(\cite{Wa10},\cite{Ma08}) and the role of amorphous structures in
the diffusion properties is still an open question
(\cite{Sm83},\cite{Wo10}). Nevertheless, the mobility of physisorbed
species is key for the evolution of the molecular complexity
\cite{Ti11}. If species like O atoms freeze out on the surface of
the grains, the chemistry is governed by H additions, leading to
numerous saturated species (H$_2$O, NH$_3$, CH$_4$, which are
chemical traps and end the chemical evolution). On the other hand,
if other atoms (O, N, C...) are mobile enough at low temperature,
other additions may open up the field of the observed molecular
complexity reached in the first stages of star formation, and that
could lead to the formation of the building blocks of life
(amino-acids). Experimentally, very few studies have already
involved physisorbed O atoms (\cite{Du10},\cite{Ji11},\cite{Wa11})
whereas theoretically, calculations exist for ordered substrates
such as graphite (\cite{Be08},\cite{La12}). On cold surfaces, ozone
reactivity has been the subject of experimental investigations for
astrochemical (\cite{Mo09},\cite{Ro11}) and atmospheric purposes
\cite{Ja10}. Formation of O$_2$ and O$_3$ on amorphous silicates
has been addressed recently \cite{Ji12} and will also be the
subject of one of our future papers. We present here experimental
evidences of tunneling of physisorbed O atoms on different substrates
(amorphous and crystalline water ice) via the study of ozone
formation, and discuss the role of the morphology of the substrate
in quantum diffusion process.

\section{Experiments and Results}

\begin{figure}[t]
\centering
\includegraphics[width=9cm]{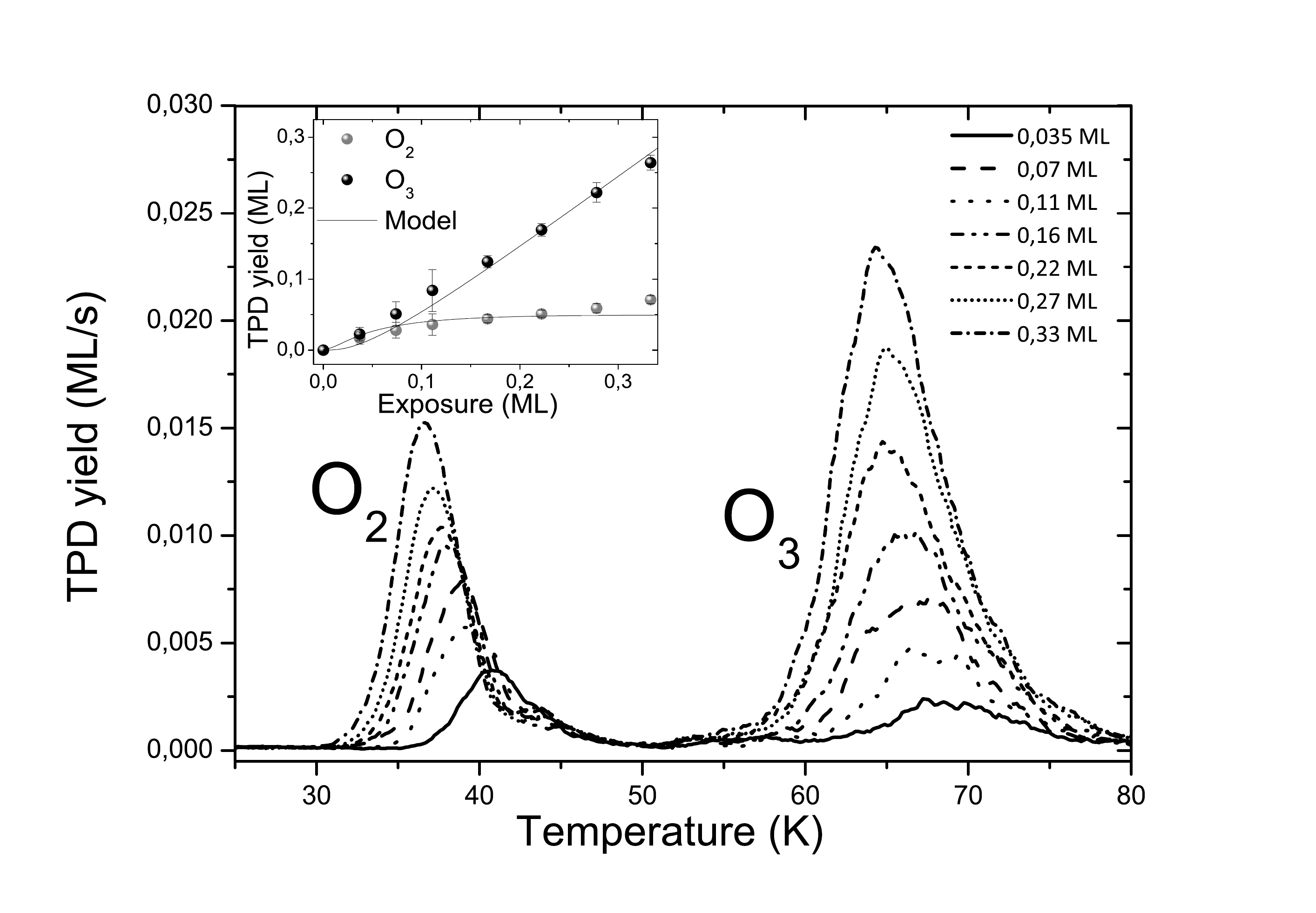}
\caption{TPD of O$_2$ and O$_3$ after
deposition of various doses (0.04 - 0.4 ML) of O/O$_2$ mixture on ASW ice held at 10~K. Inset: Areas of the TPD peaks
(in ML/s) \textit{vs} exposed O/O$_2$ dose.} \label{fig:fig2}
\end{figure}

Experiments have been performed using the FORMOLISM set-up
(described elsewhere \cite{Am07},\cite{Co12}), an ultra-high vacuum
(UHV) chamber coupled to a triply differentially pumped O beam aimed
at the temperature controlled water ice samples. O atoms are
obtained by dissociating O$_2$ gas in a microwave discharge. The
dissociation fraction $\mu$ is 71\%. It corresponds to depositing 3~O$_2$ molecules and 14~O
atoms. We have checked that atoms and molecules relax before
adsorption by scanning the kinetic energy of the ionizing electrons
of the mass spectrometer head intercepting the beam \cite{Co09}. We
detected no species with residual internal energy, i.e., having a
ionization threshold below that of the ground state.
No O$_3$ was present in the beam either.
The beam flux was calibrated using TPD (temperature-programmed-desorption) by determining the O$_2$
exposure time required to saturate the O$_2$ monolayer (ML)
\cite{No12}. In this work, the exposures are expressed in terms of
O$_2$ units, which means that 1~ML may also represent 2 layers of O
atoms or 0.66 layers of pure ozone. The compact amorphous solid
water (ASW) substrate was grown by vapor deposition on a 110K
substrate. We have also studied crystalline ice, made from an ASW
substrate annealed up to the phase transition temperature around
140K \cite{No12}, and a "porous" ice template constituted of 1.0ML
overlayer of porous water ice deposited at 10K over an ASW
substrate. This substrate has no pores, but is topologically
disordered and, particularly, presents already deep adsorption sites \cite{Fi09}.\\
In all the experiments the substrate is heated steadily (10K/min) at
the end of each deposition phase. Prior to each experiment, the sample is annealed to 100K in
order to stabilize the surface morphology before subsequent heating-cooling runs between 6.5 and 90K.\\
\figurename~\ref{fig:fig2} shows the results of experiments
performed by varying the O/O$_2$ doses. Two desorption peaks are
present: O$_2$ desorption occurs between 35K and 50K, and the
ozone desorption is observed between 55K and 75K (directly, or via
the O$_{2}^{+}$ fragments). We observe, at any coverage or
temperature, ozone formation by depositing O and O$_2$ mixtures on
ASW at 10K. O desorption is never observed. The shapes and
positions of the O$_3$ peaks are the same as those of O${_3}$
deposited from the gas phase, and are only coverage dependent.
We can thus exclude any second order desorption effects, like it should
be if O$_3$ were formed on the onset of or during desorption.
The circles in the inset of \figurename~\ref{fig:fig2} represent the
area under the TPD peaks as a function of deposited dose. The
O$_3$/O$_2$ ratio increases with coverage, O$_2$ reaches rapidly an
almost steady state while O$_3$ raises quite linearly. The solid
lines in the inset represent the model results (see below). These
experiments suggest that the reactivity of pure oxygen species is
limited to two reactions:
\begin{align}
\rm{O}+O\rightarrow  O_2  \quad & R1 \nonumber\\
\rm{O}+O_2\rightarrow  O_3  \quad & R1 \nonumber
\end{align}
The O + O$_3$ $\rightarrow$ 2O$_2$ reaction seems to be not
competitive with the others, it would not be possible to obtain
an almost pure O$_3$ sample with increasing of the coverage. The two
reactions R1 and R2 may arise from direct reactions between an
impinging atom and an adsorbed species (Eley-Rideal mechanism, ER),
or may occur by diffusion of the species on the surface
(Langmuir-Hinshelwood mechanism, LH). It is also possible that the
Hot Atom Mechanism (HAM) could have a role. Up to now HAM has been studied in conditions very different from ours: metallic surfaces, atoms with energy larger than 0.5 eV or light atoms (H or D) (\cite{To13},\cite{Di01} and references therein). The HAM is an initial
diffusion of the impinging atoms due to their residual gas phase
kinetic energy. We can consider this mechanism an extension of the
ER mechanism as it occurs during the accommodation phase. In the
case of O atoms, the kinetic energy of the beam is around 300K and
the sticking efficiency is more than 90\%, which indicates a good
transfer of energy to the water ice substrate. The binding energy is
about 1000K, thermal accommodation occurs in a few site jumps, since the atoms that stick
have lost about 300K of the total energy upon their first impact.
In what follows, we will consider the HAM mechanism included in the
ER mechanism, but with an enhanced cross section ($\times$3). The ER
mechanism happens between a gas phase reactant and a surface
reactant, it is by construction not sensitive to the surface
temperature and its efficiency depends on the coverage only. On the
contrary, the LH mechanism (as well as diffusion) depends on the
temperature of the surface. Therefore, during a TPD when temperature
increases, this mechanism could lead to the formation of other
O$_3$/O$_2$ molecules. We have attempted to check this possibility by following the evolution of O$_3$ infrared absorption band
intensity from 6.5K to 35K. Because of a high
detection limit (0.3~ML) this method could only
be applied to the highest coverage experiment in
Fig. 1, and even for this experiment the signal to
noise is too low to provide strong constraints (not
shown). Within the experimental uncertainties
the O$_3$ infrared band does not vary during the
TPD for the high coverage experiment, except
at the temperature above which ozone begins to
come off the surface ($\sim 55$K). This demonstrates
that at least some O$_3$ forms at deposition, and
the results are consistent with the theory that thermally-induced diffusion during the TPD is a
secondary effect compared to diffusion and reactions at the deposition temperature.

\begin{figure}[t]
\centering
\includegraphics[width=9cm]{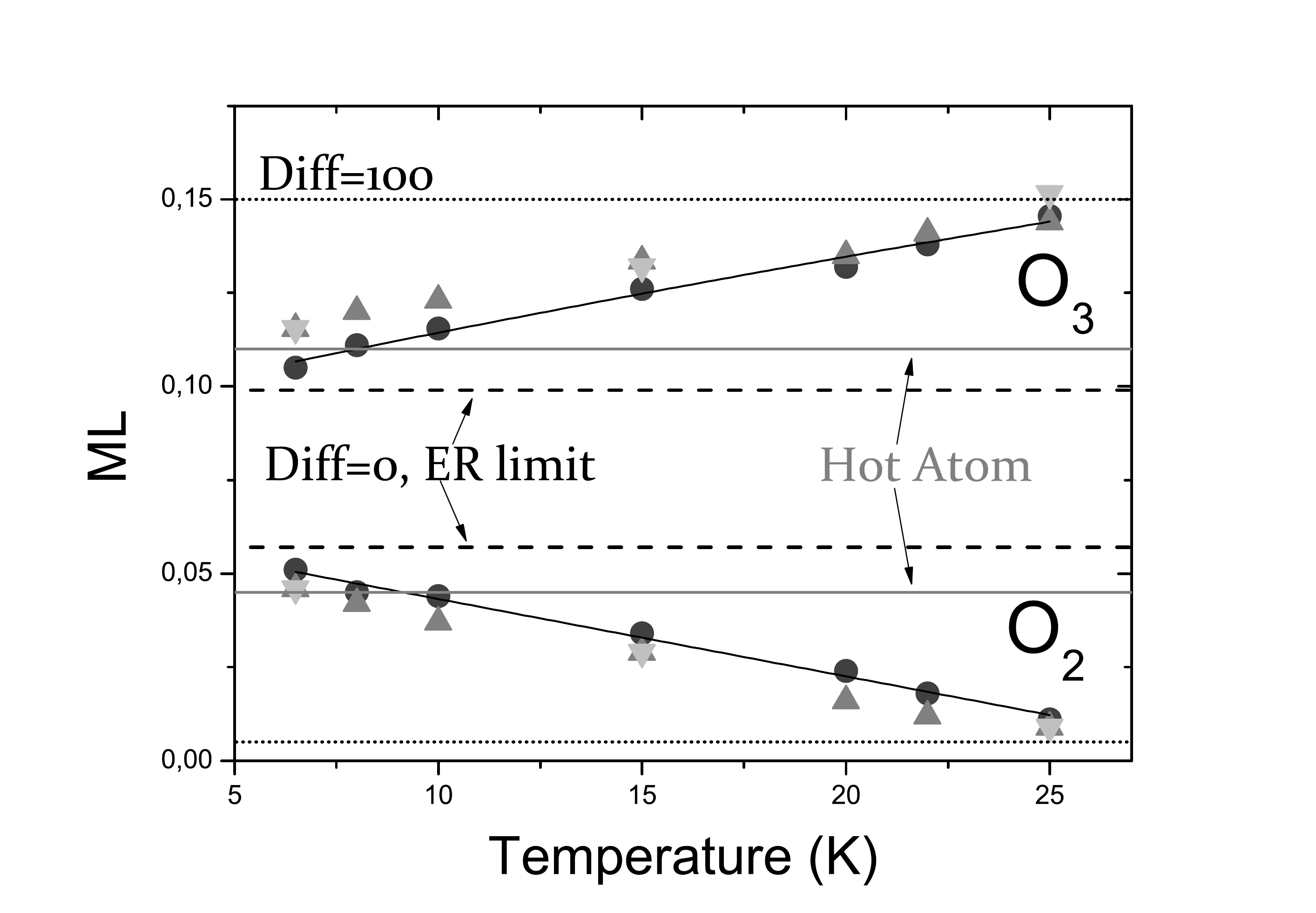}
\caption{O$_2$ and O$_3$ TPD peak areas after deposition of 0.21 ML
of O/O$_2$ mixture ($\mu =71\%$) on 3 substrates: crystalline ice
(circles), non-porous ASW ice (triangles pointing up),
and porous water ice (triangles pointing down). Solid lines:
model results. Dashed and dotted lines: O$_2$ (lower part) or O$_3$ (upper part)
amount in case of no diffusion (ER limit) and k = 100 s$^{-1}$. Dash-dotted line: model results with HAM and no diffusion.} \label{fig:fig3}
\end{figure}

To understand if diffusion effectively plays a role in the
O$_3$/O$_2$ formation, we have performed a second set of experiments
in which we varied the deposition temperature of the substrate and
the morphology of the water substrate itself, but fixing the initial
O/O$_2$ dose (coverage). As shown in \figurename~\ref{fig:fig3} we
observe that the O$_3$/O$_2$ ratio increases with the temperature of
the substrate. The evolution as a function of the exposure
temperature indicates that the temperature of the surface during
irradiation is a key parameter. In \figurename~\ref{fig:fig3} we
show the O${_3}$ and O${_2}$ yields for different temperatures and
for three different ice substrates (ASW ice, crystalline ice, and
ASW ice coated with 1ML of porous water ice). There are differences
between the three substrates, but they can be considered secondary
if compared to the global trend. Even if the temperature dependence
seems to be slow, at 25K the O$_3$ amount raises by 50\% with
respect to that at 6.5K, while in the same temperature range the
O$_2$ yield decreases by 75\%. During the TPD, the adsorbates follow
the same thermal history, and should produce the same results.
Therefore, the differences should originate at the time of the
deposition phase. ER and HAM mechanisms are
independent of the surface temperature, thus the evolution of the
O$_3$/O$_2$ ratio is due to diffusion processes.

Basically, the O$_3$/O$_2$ balance is due to the diffusion of O
atoms. If the diffusion is extremely fast, each newly adsorbed atom
scans the surface until it reacts with O$_2$ to form O$_3$. If no
O$_2$ is present, O has to wait for another O atom to form O$_2$,
which in turn will be transformed into O$_3$ via another incoming
and mobile atom. Therefore, almost all the O atoms and O$_2$
molecules are transformed into O$_3$ molecules. On the contrary, if
the diffusion is slow, one O atom has not enough time to scan the
surface to meet an O$_2$ molecule before another O comes. In this
case O-atoms accumulate until the probability for an O atom to meet
another O-atom raises, and finally O$_2$ formation is favoured.
In summary, \figurename~\ref{fig:fig2} shows that an incoming O-atom is
more likely to find O$_2$ molecules as the coverage increases. In
\figurename~\ref{fig:fig3} we show the increase of the O$_3$/O$_2$
ratio with deposition temperature which demonstrates the increase of
diffusion processes with temperature.

\section{Model and discussion}
\begin{figure}[t]
\centering
\includegraphics[width=9cm]{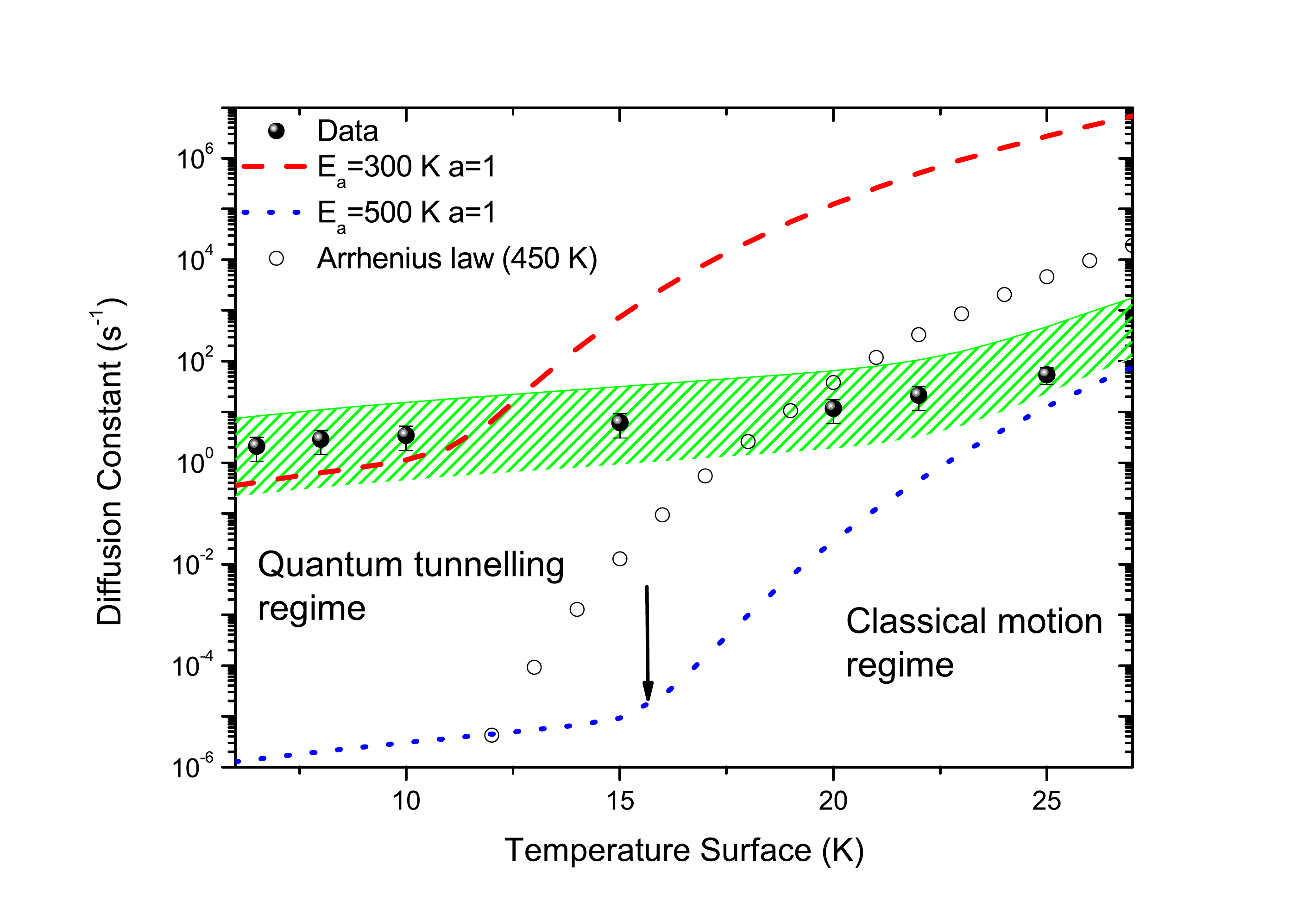}
\caption{Diffusion coefficient of O atoms on amorphous ice as function of the surface temperature. Plain circles: experimental
values. Open circles: Arrhenius law (E$_a$=450K). Dashed
and dotted line: best fit of the diffusion using both quantum and thermal
diffusion following \cite{Ji12} (a=1\AA, E$_a$=500K-300K). Pinstripe zone: a=0.7$\pm$0.05\AA, E$_a$=520K$\pm$60K. The arrow indicates the point on
the dotted curve where the quantum to classical transition occurs.}
\label{fig:fig4}
\end{figure}

As discussed in the previous section, only two reactions (R1 and R2)
occur on the surface via ER (HAM) or LH mechanisms. We have modeled
our experiments adapting a classical set of differential equations
\cite{Ka99} (see supplementary material for details). We assume a
diffusion-limited reactivity (no reaction barrier), including both
ER and LH mechanisms (same efficiency), or HAM which is estimated by
enhancing the ER mechanism, Anyway, O-atoms are of mass high enough to be able
to transfer in each single collision a relevant fraction of their kinetic
energy, and so in few jumps O-atoms are thermalized. Possible adjustments due to
thermally-induced diffusion during TPDs are also considered in order
not to exclude the possibility of incomplete reactions during the
exposure phase, even if we already noticed that it leads minor
effects. There is only one physical free parameter to adjust, $k$,
which represents the effective surface diffusion. Therefore
O$_3$/O$_2$ ratio at 25K is 45 times bigger than that at 6.5K.
Using only one adjustable parameter $k$ we reproduce perfectly all
our data sets (solid lines in \figurename~\ref{fig:fig2} and
\figurename~\ref{fig:fig3}). Other alternative scenarios (barrier to
reaction, low diffusivity, pure LH or ER mechanisms, and HAM) have been
tested without the same success. In \figurename~\ref{fig:fig3} the
boundaries of the ER mechanism and HAM are represented by two
constant lines. These limits are not sensitive to the surface
temperature. Actually, it is possible to fit the coverage dependency
with several different  hypothesis (or parameters), but it is not
possible to have both temperature and coverage dependencies
satisfied at once. Plain circles of \figurename~\ref{fig:fig4} show
the diffusion law for O as a function of the temperature, obtained
from the change in the balance of O$_2$ and O$_3$ production,
assuming a diffusion dominated process. The diffusion coefficient
increases by a factor of 50, but the logarithm scaling of the figure
tends to flatten this aspect of the experimental results. The trend
is somehow surprising because the measured diffusion barrier does not
follow an Arrhenius law. Empty circles in \figurename~\ref{fig:fig4}
represent a typical Arrhenius law with a diffusion barrier of 450K.
Therefore, a pure thermal diffusion does not represent well our
data. The diffusion is better simulated using the quantum tunneling
of a square barrier, as described in Messiah's book \cite{Me73}.
We use two physical parameters:  the width $a$ and the height of the
barrier $E_a$. As already described \cite{Ca04}, diffusion
includes two components, quantum tunneling that dominates at low
temperatures, and thermal diffusion predominant at higher
temperatures. The arrow in \figurename~\ref{fig:fig4} represents the
critical temperature where the transition occurs. In our
experimental data we observe that the quantum to classical regime
occurs at around 20K. We find the best results for $a$=0.70$\pm$
0.05$\text{\AA}$ and $E_a$=520$\pm$60K. These two parameters have
different effects on the diffusion curves as shown in
\figurename~\ref{fig:fig4} (dashed and dotted lines). The pinstriped region in
\figurename~\ref{fig:fig4} represents the validity zone of the
solution we found, since it is possible to partly compensate for the
variation of the height of the barrier by changing its width.
Adopting a square barrier is an extreme assumption, and the apparent
low value of the height we found may be due to this unrealistic
shape.

The main results of our study are the following: first, O-atom
diffusion is governed by quantum tunneling up to 20K. Our
experiments show nicely the transition from the quantum world to the
classic world. The thickness of the barrier (0.7$\text{\AA}$) may
also be due to the amorphous nature of the ice.
Indeed, one can consider that the diffusion on amorphous surface is dominated by the
fastest jumps between adsorption sites, and that the diffusion
barriers limiting the diffusion are the weakest ones \cite{Ka12}.
Therefore, the apparent low value of width is probably an estimation
of the lower limit of the distribution of barrier widths of this
disordered substrate. 
We want now discuss the signification of a width barrier of 0.7$\text{\AA}$. On water ice the mean distance of two adsorption sites is 3$\text{\AA}$, while the de Broglie wavelength($\lambda_b$) associated to O atoms varies from 2.8$\text{\AA}$ at 6.5K to 1.5$\text{\AA}$ at 20K. It corresponds to the typical size where quantum effects are important. We can compare the 3$\text{\AA}$ mean distance with the double (2 atoms diffusion-reaction) of $\lambda_b$. For temperatures lower than 22K, quantum effects should thus be important. This is actually what we observe. The slightly lower value (0.7 instead of 1$\text{\AA}$ guessed in the literature), could also be due to the disordered nature of the surface that distribute distances between adjacent adsorption sites, and so reduces some of them. Therefore cold O atoms would easily overlap 2 neighbouring sites.\\
Secondly, comparing the different morphologies
of the ice allows us to conclude that the topological disorder of
the substrate does not deeply affect the diffusion regime. This probably means that the lower limits of the distributions of barrier
widths are not too different from each other for the substates considered
here. However, we find that the diffusion on the crystalline surface is faster than
that on amorphous ice. This could be explained as follows: i) the wave packet describing the
adsorbed atom diffuses more quickly in the periodic potential of a
crystalline surface than in the amorphous one \cite{Ka12}. ii) Another study \cite{Wo10} showed
that the diffusion and reactive properties are greatly changed by
the occurrence of deep and shallow sites. In our case, we find that
the presence of deep sites (``porous'' substrate) does not reduce
significantly the effective diffusion. The trapping sites may force
one O atom to stay in one adsorption site, but they cannot prevent
another atom from reaching it, especially if tunneling dominates.\\
Third conclusion, we find that the diffusion of oxygen at 10K is
$k=5 s^{-1}$, which corresponds in classical Arrhenius formalism to
thermal diffusion barrier of 300K. The diffusion is faster than
previously guessed by astronomers (diffusion barrier of 400K
\cite{Ti82} or even  900K \cite{Ca10}. However, it is not very
far from that of O diffusion in matrices, 240$\pm$80K
\cite{Be96}, which could be considered an upper value for surface
diffusion. Based upon arguments of polarizability scaling
\cite{Ti11}, and the mean value of \cite{Ma08}, we would expect a
value of 275$\pm$30K, which is consistent with what is derived
here. From our model we can indirectly conclude that the maximal
barrier height for the O$_2$+O reaction is 190K. Actually, if the
barriers were higher, there would be noticeable discrepancies in the
O$_2$/O$_3$ ratio, which we do not observe. It is probably even
smaller as estimated in a previous study \cite{Be96}. Implications
for solid state astrochemistry are of major importance. It was
usually thought that the chemistry was mostly driven by H diffusion,
and therefore final products were mostly hydrogen saturated species
such as H$_2$O, NH$_3$, CH$_4$, CH$_3$OH. We can affirm now that O
addition chemistry is competitive with H additions, because of the
comparable budget of O atoms and H atoms in dense and UV protected
interstellar environments \cite{Ca02}. Furthermore, if we scale the
diffusion for other atoms such as C and N, on the basis of their
polarizability alone - the main parameter for physisorption - there
is a reasonable range of temperatures ($\le$15K) where mobility of O,
C and N is activated. For this reason, the production of O, C, and N
bearing molecules can grow, avoiding saturated chemical traps. This
could also be one of the source of complex non-volatile organic
compounds observed in meteorites, such as amino-acids.

\begin{acknowledgments}
We acknowledge the support of the
national PCMI programme founded by CNRS, the Conseil Regional d'Ile
de France through SESAME programmes (contract I-07–597R). MM
acknowledges LASSIE, a European FP7 ITN Community's Seventh Framework
Programme under Grant Agreement No. 238258. MA thanks COST ACTION
CM0805.We also thank the unknown referees for the fruitful comments and suggestions that helped make a better paper.
\end{acknowledgments}

\section{Supplementary Materials: Model. Set of differential equations used to analyze the experimental data.}
For the sake of clarity, the equation are detailed in two sets, even
if they are solved simultaneously. One corresponds to the diffusion
(LH mechanism) and another to the ER mechanism. The LH mechanism is described as follows
\begin{align}
\frac{dO}{dt} =& 4kOO-kOO_2 \nonumber\\
\frac{dO_2}{dt} =& 2kOO-kOO_2 \nonumber\\
\frac{dO_3}{dt} =& kOO_2 \nonumber
\end{align}
O$_x$ is the fraction of occupied sites of the species \textit{x}
(it is equal to one when the coverage is 1~ML=10$^{15}$cm$^{-2}$),
and \textit{k} is the diffusion parameter (in s$^{-1}$). The increase of the O$_3$ population is
proportional to the density of reactants times the diffusion
coefficient. Eley-Rideal mechanism is described as follows:
\begin{align}
\frac{dO}{dt} =& 2\mu \phi (1-2O-O_2)-(1-\mu)\phi O \nonumber\\
\frac{dO_2}{dt} =& (1-\mu)\phi (1-O) - 2\mu \phi (O_2-O)\nonumber\\
\frac{dO_3}{dt} =& (1-\mu)\phi O + 2\mu \phi O_2 \nonumber
\end{align}
$\phi$ is the flux and $\mu$ is the dissociation rate. The cross
section is supposed to be as large as the adsorption site area (so
equal to 1, not shown), which can be considered as an upper value.
Nevertheless, the ER mechanism does not take a large part in the
calculations, especially at low coverages. The time is integrated
for the duration of the experimental exposure.

\bibliography{apssamp}% Produces the bibliography via BibTeX.

\providecommand{\noopsort}[1]{}\providecommand{\singleletter}[1]{#1}%
\begin{thebibliography}{34}%
\makeatletter
\providecommand \@ifxundefined [1]{%
 \@ifx{#1\undefined}
}%
\providecommand \@ifnum [1]{%
 \ifnum #1\expandafter \@firstoftwo
 \else \expandafter \@secondoftwo
 \fi
}%
\providecommand \@ifx [1]{%
 \ifx #1\expandafter \@firstoftwo
 \else \expandafter \@secondoftwo
 \fi
}%
\providecommand \natexlab [1]{#1}%
\providecommand \enquote  [1]{``#1''}%
\providecommand \bibnamefont  [1]{#1}%
\providecommand \bibfnamefont [1]{#1}%
\providecommand \citenamefont [1]{#1}%
\providecommand \href@noop [0]{\@secondoftwo}%
\providecommand \href [0]{\begingroup \@sanitize@url \@href}%
\providecommand \@href[1]{\@@startlink{#1}\@@href}%
\providecommand \@@href[1]{\endgroup#1\@@endlink}%
\providecommand \@sanitize@url [0]{\catcode `\\12\catcode `\$12\catcode
  `\&12\catcode `\#12\catcode `\^12\catcode `\_12\catcode `\%12\relax}%
\providecommand \@@startlink[1]{}%
\providecommand \@@endlink[0]{}%
\providecommand \url  [0]{\begingroup\@sanitize@url \@url }%
\providecommand \@url [1]{\endgroup\@href {#1}{\urlprefix }}%
\providecommand \urlprefix  [0]{URL }%
\providecommand \Eprint [0]{\href }%
\providecommand \doibase [0]{http://dx.doi.org/}%
\providecommand \selectlanguage [0]{\@gobble}%
\providecommand \bibinfo  [0]{\@secondoftwo}%
\providecommand \bibfield  [0]{\@secondoftwo}%
\providecommand \translation [1]{[#1]}%
\providecommand \BibitemOpen [0]{}%
\providecommand \bibitemStop [0]{}%
\providecommand \bibitemNoStop [0]{.\EOS\space}%
\providecommand \EOS [0]{\spacefactor3000\relax}%
\providecommand \BibitemShut  [1]{\csname bibitem#1\endcsname}%
\let\auto@bib@innerbib\@empty
%</preamble>
\bibitem [{\citenamefont {Amiaud}\ \emph {et~al.}(2007)\citenamefont {Amiaud},
  \citenamefont {Dulieu}, \citenamefont {Fillion}, \citenamefont {Momeni},\
  and\ \citenamefont {J.~L.~Lemaire}}]{Am07}%
  \BibitemOpen
  \bibfield  {author} {\bibinfo {author} {\bibfnamefont {L.}~\bibnamefont
  {Amiaud}}, \bibinfo {author} {\bibfnamefont {F.}~\bibnamefont {Dulieu}},
  \bibinfo {author} {\bibfnamefont {J.-H.}\ \bibnamefont {Fillion}}, \bibinfo
  {author} {\bibfnamefont {A.}~\bibnamefont {Momeni}}, \ and\ \bibinfo {author}
  {\bibfnamefont {J.~L.}\ \bibnamefont {J.~L.~Lemaire}},\ }\href@noop {}
  {\bibfield  {journal} {\bibinfo  {journal} {JCP}\ }\textbf {\bibinfo {volume}
  {127}},\ \bibinfo {pages} {094702} (\bibinfo {year} {2007})}\BibitemShut
  {NoStop}%
\bibitem [{\citenamefont {Benderskii}\ and\ \citenamefont
  {Wight}(1996)}]{Be96}%
  \BibitemOpen
  \bibfield  {author} {\bibinfo {author} {\bibfnamefont {A.~V.}\ \bibnamefont
  {Benderskii}}\ and\ \bibinfo {author} {\bibfnamefont {C.~A.}\ \bibnamefont
  {Wight}},\ }\href@noop {} {\bibfield  {journal} {\bibinfo  {journal} {JCP}\
  }\textbf {\bibinfo {volume} {104}},\ \bibinfo {pages} {85} (\bibinfo {year}
  {1996})}\BibitemShut {NoStop}%
\bibitem [{\citenamefont {Bergeron}\ \emph {et~al.}(2008)\citenamefont
  {Bergeron}, \citenamefont {Rougeau}, \citenamefont {Sidis}, \citenamefont
  {Sizun}, \citenamefont {Teillet-Billy},\ and\ \citenamefont
  {Aguillon}}]{Be08}%
  \BibitemOpen
  \bibfield  {author} {\bibinfo {author} {\bibfnamefont {H.}~\bibnamefont
  {Bergeron}}, \bibinfo {author} {\bibfnamefont {N.}~\bibnamefont {Rougeau}},
  \bibinfo {author} {\bibfnamefont {V.}~\bibnamefont {Sidis}}, \bibinfo
  {author} {\bibfnamefont {M.}~\bibnamefont {Sizun}}, \bibinfo {author}
  {\bibfnamefont {D.}~\bibnamefont {Teillet-Billy}}, \ and\ \bibinfo {author}
  {\bibfnamefont {F.}~\bibnamefont {Aguillon}},\ }\href@noop {} {\bibfield
  {journal} {\bibinfo  {journal} {JCP A}\ }\textbf {\bibinfo {volume} {112}},\
  \bibinfo {pages} {11921} (\bibinfo {year} {2008})}\BibitemShut {NoStop}%
\bibitem [{\citenamefont {Caselli}\ \emph {et~al.}(2002)\citenamefont
  {Caselli}, \citenamefont {Stantcheva}, \citenamefont {Shematovich},\ and\
  \citenamefont {Herbst}}]{Ca02}%
  \BibitemOpen
  \bibfield  {author} {\bibinfo {author} {\bibfnamefont {P.}~\bibnamefont
  {Caselli}}, \bibinfo {author} {\bibfnamefont {O.}~\bibnamefont {Stantcheva},
  \bibfnamefont {T.~Shalabiea}}, \bibinfo {author} {\bibfnamefont
  {V.}~\bibnamefont {Shematovich}}, \ and\ \bibinfo {author} {\bibfnamefont
  {E.}~\bibnamefont {Herbst}},\ }\href@noop {} {\bibfield  {journal} {\bibinfo
  {journal} {P\&SS}\ }\textbf {\bibinfo {volume} {50}},\ \bibinfo {pages}
  {1257C} (\bibinfo {year} {2002})}\BibitemShut {NoStop}%
\bibitem [{\citenamefont {Cazaux}\ \emph {et~al.}(522)\citenamefont {Cazaux},
  \citenamefont {Cobut}, \citenamefont {Marseille}, \citenamefont {Spaans},\
  and\ \citenamefont {Caselli}}]{Ca10}%
  \BibitemOpen
  \bibfield  {author} {\bibinfo {author} {\bibfnamefont {S.}~\bibnamefont
  {Cazaux}}, \bibinfo {author} {\bibfnamefont {V.}~\bibnamefont {Cobut}},
  \bibinfo {author} {\bibfnamefont {M.}~\bibnamefont {Marseille}}, \bibinfo
  {author} {\bibfnamefont {M.}~\bibnamefont {Spaans}}, \ and\ \bibinfo {author}
  {\bibfnamefont {P.}~\bibnamefont {Caselli}},\ }\href@noop {} {\bibfield
  {journal} {\bibinfo  {journal} {A\&A}\ }\textbf {\bibinfo {volume} {A74}},\
  \bibinfo {pages} {2010} (\bibinfo {year} {522})}\BibitemShut {NoStop}%
\bibitem [{\citenamefont {Cazaux}\ and\ \citenamefont {Tielens}(2004)}]{Ca04}%
  \BibitemOpen
  \bibfield  {author} {\bibinfo {author} {\bibfnamefont {S.}~\bibnamefont
  {Cazaux}}\ and\ \bibinfo {author} {\bibfnamefont {A.}~\bibnamefont
  {Tielens}},\ }\href@noop {} {\bibfield  {journal} {\bibinfo  {journal} {ApJ}\
  }\textbf {\bibinfo {volume} {604}},\ \bibinfo {pages} {222} (\bibinfo {year}
  {2004})}\BibitemShut {NoStop}%
\bibitem [{\citenamefont {Congiu}\ \emph {et~al.}(2012)\citenamefont {Congiu},
  \citenamefont {Chaabouni}, \citenamefont {Laffon}, \citenamefont {Parent},
  \citenamefont {Baouche},\ and\ \citenamefont {Dulieu}}]{Co12}%
  \BibitemOpen
  \bibfield  {author} {\bibinfo {author} {\bibfnamefont {E.}~\bibnamefont
  {Congiu}}, \bibinfo {author} {\bibfnamefont {H.}~\bibnamefont {Chaabouni}},
  \bibinfo {author} {\bibfnamefont {C.}~\bibnamefont {Laffon}}, \bibinfo
  {author} {\bibfnamefont {P.}~\bibnamefont {Parent}}, \bibinfo {author}
  {\bibfnamefont {S.}~\bibnamefont {Baouche}}, \ and\ \bibinfo {author}
  {\bibfnamefont {F.}~\bibnamefont {Dulieu}},\ }\href@noop {} {\bibfield
  {journal} {\bibinfo  {journal} {JCP}\ }\textbf {\bibinfo {volume} {137}},\
  \bibinfo {pages} {054713} (\bibinfo {year} {2012})}\BibitemShut {NoStop}%
\bibitem [{\citenamefont {Congiu}\ \emph {et~al.}(2009)\citenamefont {Congiu},
  \citenamefont {Matar}, \citenamefont {Kristensen}, \citenamefont {Dulieu},\
  and\ \citenamefont {Lemaire}}]{Co09}%
  \BibitemOpen
  \bibfield  {author} {\bibinfo {author} {\bibfnamefont {E.}~\bibnamefont
  {Congiu}}, \bibinfo {author} {\bibfnamefont {E.}~\bibnamefont {Matar}},
  \bibinfo {author} {\bibfnamefont {L.~E.}\ \bibnamefont {Kristensen}},
  \bibinfo {author} {\bibfnamefont {F.}~\bibnamefont {Dulieu}}, \ and\ \bibinfo
  {author} {\bibfnamefont {J.~L.}\ \bibnamefont {Lemaire}},\ }\href@noop {}
  {\bibfield  {journal} {\bibinfo  {journal} {MNRAS}\ }\textbf {\bibinfo
  {volume} {397L}},\ \bibinfo {pages} {96C} (\bibinfo {year}
  {2009})}\BibitemShut {NoStop}%
\bibitem [{\citenamefont {Cuppen}\ and\ \citenamefont {Herbst}(2007)}]{Cu07}%
  \BibitemOpen
  \bibfield  {author} {\bibinfo {author} {\bibfnamefont {H.~M.}\ \bibnamefont
  {Cuppen}}\ and\ \bibinfo {author} {\bibfnamefont {E.}~\bibnamefont
  {Herbst}},\ }\href@noop {} {\bibfield  {journal} {\bibinfo  {journal} {ApJ}\
  }\textbf {\bibinfo {volume} {668}},\ \bibinfo {pages} {294} (\bibinfo {year}
  {2007})}\BibitemShut {NoStop}%
\bibitem [{\citenamefont {Dinger}\ and\ \citenamefont {Kuppers}(2001)}]{Di01}%
  \BibitemOpen
  \bibfield  {author} {\bibinfo {author} {\bibfnamefont {L.~C.}\ \bibnamefont
  {Dinger}, \bibfnamefont {A.}}\ and\ \bibinfo {author} {\bibfnamefont
  {J.}~\bibnamefont {Kuppers}},\ }\href@noop {} {\bibfield  {journal} {\bibinfo
   {journal} {J. Chem. Phys.}\ }\textbf {\bibinfo {volume} {114}},\ \bibinfo
  {pages} {12, 5338} (\bibinfo {year} {2001})}\BibitemShut {NoStop}%
\bibitem [{\citenamefont {Dulieu}\ \emph {et~al.}(2010)\citenamefont {Dulieu},
  \citenamefont {Amiaud}, \citenamefont {Congiu}, \citenamefont {Fillion},
  \citenamefont {Matar}, \citenamefont {Momeni}, \citenamefont {Pirronello},\
  and\ \citenamefont {Lemaire}}]{Du10}%
  \BibitemOpen
  \bibfield  {author} {\bibinfo {author} {\bibfnamefont {F.}~\bibnamefont
  {Dulieu}}, \bibinfo {author} {\bibfnamefont {L.}~\bibnamefont {Amiaud}},
  \bibinfo {author} {\bibfnamefont {E.}~\bibnamefont {Congiu}}, \bibinfo
  {author} {\bibfnamefont {J.-H.}\ \bibnamefont {Fillion}}, \bibinfo {author}
  {\bibfnamefont {E.}~\bibnamefont {Matar}}, \bibinfo {author} {\bibfnamefont
  {A.}~\bibnamefont {Momeni}}, \bibinfo {author} {\bibfnamefont
  {V.}~\bibnamefont {Pirronello}}, \ and\ \bibinfo {author} {\bibfnamefont
  {J.~L.}\ \bibnamefont {Lemaire}},\ }\href@noop {} {\bibfield  {journal}
  {\bibinfo  {journal} {A\&A}\ }\textbf {\bibinfo {volume} {512}},\ \bibinfo
  {pages} {A30} (\bibinfo {year} {2010})}\BibitemShut {NoStop}%
\bibitem [{\citenamefont {Ehrlich}(1994)}]{Eh94}%
  \BibitemOpen
  \bibfield  {author} {\bibinfo {author} {\bibfnamefont {G.}~\bibnamefont
  {Ehrlich}},\ }\href@noop {} {\bibfield  {journal} {\bibinfo  {journal} {Surf.
  Sci.}\ }\textbf {\bibinfo {volume} {299}},\ \bibinfo {pages} {628} (\bibinfo
  {year} {1994})}\BibitemShut {NoStop}%
\bibitem [{\citenamefont {Fillion}\ \emph {et~al.}(2009)\citenamefont
  {Fillion}, \citenamefont {Amiaud}, \citenamefont {Congiu}, \citenamefont
  {Dulieu}, \citenamefont {Momeni},\ and\ \citenamefont {Lemaire}}]{Fi09}%
  \BibitemOpen
  \bibfield  {author} {\bibinfo {author} {\bibfnamefont {J.-H.}\ \bibnamefont
  {Fillion}}, \bibinfo {author} {\bibfnamefont {L.}~\bibnamefont {Amiaud}},
  \bibinfo {author} {\bibfnamefont {E.}~\bibnamefont {Congiu}}, \bibinfo
  {author} {\bibfnamefont {F.}~\bibnamefont {Dulieu}}, \bibinfo {author}
  {\bibfnamefont {A.}~\bibnamefont {Momeni}}, \ and\ \bibinfo {author}
  {\bibfnamefont {J.-L.}\ \bibnamefont {Lemaire}},\ }\href@noop {} {\bibfield
  {journal} {\bibinfo  {journal} {PCCP}\ }\textbf {\bibinfo {volume} {11}},\
  \bibinfo {pages} {4396} (\bibinfo {year} {2009})}\BibitemShut {NoStop}%
\bibitem [{\citenamefont {Gillan}(1987)}]{Gi87}%
  \BibitemOpen
  \bibfield  {author} {\bibinfo {author} {\bibfnamefont {M.~J.}\ \bibnamefont
  {Gillan}},\ }\href@noop {} {\bibfield  {journal} {\bibinfo  {journal} {JP C:
  Solid State Phys.}\ }\textbf {\bibinfo {volume} {20}},\ \bibinfo {pages}
  {3621} (\bibinfo {year} {1987})}\BibitemShut {NoStop}%
\bibitem [{\citenamefont {Goumans}\ and\ \citenamefont
  {Andersson}(2010)}]{Go10}%
  \BibitemOpen
  \bibfield  {author} {\bibinfo {author} {\bibfnamefont {T.~P.~M.}\
  \bibnamefont {Goumans}}\ and\ \bibinfo {author} {\bibfnamefont
  {S.}~\bibnamefont {Andersson}},\ }\href@noop {} {\bibfield  {journal}
  {\bibinfo  {journal} {MNRAS}\ }\textbf {\bibinfo {volume} {406}},\ \bibinfo
  {pages} {2213} (\bibinfo {year} {2010})}\BibitemShut {NoStop}%
\bibitem [{\citenamefont {Janssen}\ and\ \citenamefont {Tuzson}(2010)}]{Ja10}%
  \BibitemOpen
  \bibfield  {author} {\bibinfo {author} {\bibfnamefont {C.}~\bibnamefont
  {Janssen}}\ and\ \bibinfo {author} {\bibfnamefont {B.}~\bibnamefont
  {Tuzson}},\ }\href@noop {} {\bibfield  {journal} {\bibinfo  {journal} {JPC
  A}\ }\textbf {\bibinfo {volume} {114}},\ \bibinfo {pages} {9709} (\bibinfo
  {year} {2010})}\BibitemShut {NoStop}%
\bibitem [{\citenamefont {Jing}\ \emph {et~al.}(2011)\citenamefont {Jing},
  \citenamefont {He}, \citenamefont {Brucato}, \citenamefont {De~Sio},
  \citenamefont {Tozzetti},\ and\ \citenamefont {Vidali}}]{Ji11}%
  \BibitemOpen
  \bibfield  {author} {\bibinfo {author} {\bibfnamefont {D.}~\bibnamefont
  {Jing}}, \bibinfo {author} {\bibfnamefont {J.}~\bibnamefont {He}}, \bibinfo
  {author} {\bibfnamefont {J.}~\bibnamefont {Brucato}}, \bibinfo {author}
  {\bibfnamefont {A.}~\bibnamefont {De~Sio}}, \bibinfo {author} {\bibfnamefont
  {L.}~\bibnamefont {Tozzetti}}, \ and\ \bibinfo {author} {\bibfnamefont
  {G.}~\bibnamefont {Vidali}},\ }\href@noop {} {\bibfield  {journal} {\bibinfo
  {journal} {ApJ}\ }\textbf {\bibinfo {volume} {714L}},\ \bibinfo {pages} {9J}
  (\bibinfo {year} {2011})}\BibitemShut {NoStop}%
\bibitem [{\citenamefont {Jing}\ \emph {et~al.}(2012)\citenamefont {Jing},
  \citenamefont {He}, \citenamefont {Brucato}, \citenamefont {Vidali},
  \citenamefont {Tozzetti},\ and\ \citenamefont {De~Sio}}]{Ji12}%
  \BibitemOpen
  \bibfield  {author} {\bibinfo {author} {\bibfnamefont {D.}~\bibnamefont
  {Jing}}, \bibinfo {author} {\bibfnamefont {J.}~\bibnamefont {He}}, \bibinfo
  {author} {\bibfnamefont {J.~R.}\ \bibnamefont {Brucato}}, \bibinfo {author}
  {\bibfnamefont {G.}~\bibnamefont {Vidali}}, \bibinfo {author} {\bibfnamefont
  {L.}~\bibnamefont {Tozzetti}}, \ and\ \bibinfo {author} {\bibfnamefont
  {A.}~\bibnamefont {De~Sio}},\ }\href@noop {} {\bibfield  {journal} {\bibinfo
  {journal} {ApJ}\ }\textbf {\bibinfo {volume} {756}},\ \bibinfo {pages} {98}
  (\bibinfo {year} {2012})}\BibitemShut {NoStop}%
\bibitem [{\citenamefont {Karssemeijer}\ \emph {et~al.}(2012)\citenamefont
  {Karssemeijer}, \citenamefont {Pedersen}, \citenamefont {Jonsson},\ and\
  \citenamefont {Cuppen}}]{Ka12}%
  \BibitemOpen
  \bibfield  {author} {\bibinfo {author} {\bibfnamefont {L.}~\bibnamefont
  {Karssemeijer}}, \bibinfo {author} {\bibfnamefont {A.}~\bibnamefont
  {Pedersen}}, \bibinfo {author} {\bibfnamefont {H.}~\bibnamefont {Jonsson}}, \
  and\ \bibinfo {author} {\bibfnamefont {H.}~\bibnamefont {Cuppen}},\
  }\href@noop {} {\bibfield  {journal} {\bibinfo  {journal} {PCCP}\ }\textbf
  {\bibinfo {volume} {134}},\ \bibinfo {pages} {084504} (\bibinfo {year}
  {2012})}\BibitemShut {NoStop}%
\bibitem [{\citenamefont {Katz}\ \emph {et~al.}(1999)\citenamefont {Katz},
  \citenamefont {Furman}, \citenamefont {Biham}, \citenamefont {Pirronello},\
  and\ \citenamefont {Vidali}}]{Ka99}%
  \BibitemOpen
  \bibfield  {author} {\bibinfo {author} {\bibfnamefont {N.}~\bibnamefont
  {Katz}}, \bibinfo {author} {\bibfnamefont {I.}~\bibnamefont {Furman}},
  \bibinfo {author} {\bibfnamefont {O.}~\bibnamefont {Biham}}, \bibinfo
  {author} {\bibfnamefont {V.}~\bibnamefont {Pirronello}}, \ and\ \bibinfo
  {author} {\bibfnamefont {G.}~\bibnamefont {Vidali}},\ }\href@noop {}
  {\bibfield  {journal} {\bibinfo  {journal} {ApJ}\ }\textbf {\bibinfo {volume}
  {522}},\ \bibinfo {pages} {305} (\bibinfo {year} {1999})}\BibitemShut
  {NoStop}%
\bibitem [{\citenamefont {Larciprete}\ \emph {et~al.}(2012)\citenamefont
  {Larciprete}, \citenamefont {Lacovig}, \citenamefont {Gardonio},
  \citenamefont {Baraldi},\ and\ \citenamefont {Lizzit}}]{La12}%
  \BibitemOpen
  \bibfield  {author} {\bibinfo {author} {\bibfnamefont {R.}~\bibnamefont
  {Larciprete}}, \bibinfo {author} {\bibfnamefont {P.}~\bibnamefont {Lacovig}},
  \bibinfo {author} {\bibfnamefont {S.}~\bibnamefont {Gardonio}}, \bibinfo
  {author} {\bibfnamefont {A.}~\bibnamefont {Baraldi}}, \ and\ \bibinfo
  {author} {\bibfnamefont {S.}~\bibnamefont {Lizzit}},\ }\href@noop {}
  {\bibfield  {journal} {\bibinfo  {journal} {JPC C}\ }\textbf {\bibinfo
  {volume} {116}},\ \bibinfo {pages} {9900} (\bibinfo {year}
  {2012})}\BibitemShut {NoStop}%
\bibitem [{\citenamefont {Matar}\ \emph {et~al.}(2008)\citenamefont {Matar},
  \citenamefont {Congiu}, \citenamefont {Dulieu}, \citenamefont {Momeni},\ and\
  \citenamefont {Lemaire}}]{Ma08}%
  \BibitemOpen
  \bibfield  {author} {\bibinfo {author} {\bibfnamefont {E.}~\bibnamefont
  {Matar}}, \bibinfo {author} {\bibfnamefont {E.}~\bibnamefont {Congiu}},
  \bibinfo {author} {\bibfnamefont {F.}~\bibnamefont {Dulieu}}, \bibinfo
  {author} {\bibfnamefont {A.}~\bibnamefont {Momeni}}, \ and\ \bibinfo {author}
  {\bibfnamefont {J.~L.}\ \bibnamefont {Lemaire}},\ }\href@noop {} {\bibfield
  {journal} {\bibinfo  {journal} {A\&A}\ }\textbf {\bibinfo {volume} {492}},\
  \bibinfo {pages} {L17} (\bibinfo {year} {2008})}\BibitemShut {NoStop}%
\bibitem [{\citenamefont {Messiah}(1973)}]{Me73}%
  \BibitemOpen
  \bibfield  {author} {\bibinfo {author} {\bibfnamefont {A.}~\bibnamefont
  {Messiah}},\ }\href@noop {} {\emph {\bibinfo {title} {Quantum Mechanics}}}\
  (\bibinfo  {publisher} {Wiley Amsterdam, North Holland,},\ \bibinfo {year}
  {1973})\BibitemShut {NoStop}%
\bibitem [{\citenamefont {Mokrane}\ \emph {et~al.}(2009)\citenamefont
  {Mokrane}, \citenamefont {Chaabouni}, \citenamefont {Accolla}, \citenamefont
  {Dulieu}, \citenamefont {Chehrouri},\ and\ \citenamefont {Lemaire}}]{Mo09}%
  \BibitemOpen
  \bibfield  {author} {\bibinfo {author} {\bibfnamefont {H.}~\bibnamefont
  {Mokrane}}, \bibinfo {author} {\bibfnamefont {H.}~\bibnamefont {Chaabouni}},
  \bibinfo {author} {\bibfnamefont {E.}~\bibnamefont {Accolla}, \bibfnamefont
  {M.~andCongiu}}, \bibinfo {author} {\bibfnamefont {F.}~\bibnamefont
  {Dulieu}}, \bibinfo {author} {\bibfnamefont {M.}~\bibnamefont {Chehrouri}}, \
  and\ \bibinfo {author} {\bibfnamefont {J.~L.}\ \bibnamefont {Lemaire}},\
  }\href@noop {} {\bibfield  {journal} {\bibinfo  {journal} {ApJL}\ }\textbf
  {\bibinfo {volume} {705}},\ \bibinfo {pages} {L195} (\bibinfo {year}
  {2009})}\BibitemShut {NoStop}%
\bibitem [{\citenamefont {Noble}\ \emph {et~al.}(2012)\citenamefont {Noble},
  \citenamefont {Congiu}, \citenamefont {Dulieu},\ and\ \citenamefont
  {Fraser}}]{No12}%
  \BibitemOpen
  \bibfield  {author} {\bibinfo {author} {\bibfnamefont {J.~A.}\ \bibnamefont
  {Noble}}, \bibinfo {author} {\bibfnamefont {E.}~\bibnamefont {Congiu}},
  \bibinfo {author} {\bibfnamefont {F.}~\bibnamefont {Dulieu}}, \ and\ \bibinfo
  {author} {\bibfnamefont {H.~J.}\ \bibnamefont {Fraser}},\ }\href@noop {}
  {\bibfield  {journal} {\bibinfo  {journal} {MNRAS}\ }\textbf {\bibinfo
  {volume} {421}},\ \bibinfo {pages} {768–779} (\bibinfo {year}
  {2012})}\BibitemShut {NoStop}%
\bibitem [{\citenamefont {Pagani}\ \emph {et~al.}(2007)\citenamefont {Pagani},
  \citenamefont {Bacmann}, \citenamefont {Cabrit},\ and\ \citenamefont
  {Vastel}}]{Pa07}%
  \BibitemOpen
  \bibfield  {author} {\bibinfo {author} {\bibfnamefont {L.}~\bibnamefont
  {Pagani}}, \bibinfo {author} {\bibfnamefont {A.}~\bibnamefont {Bacmann}},
  \bibinfo {author} {\bibfnamefont {S.}~\bibnamefont {Cabrit}}, \ and\ \bibinfo
  {author} {\bibfnamefont {C.}~\bibnamefont {Vastel}},\ }\href@noop {}
  {\bibfield  {journal} {\bibinfo  {journal} {A\&A}\ }\textbf {\bibinfo
  {volume} {467}},\ \bibinfo {pages} {179} (\bibinfo {year}
  {2007})}\BibitemShut {NoStop}%
\bibitem [{\citenamefont {Romanzin}\ \emph {et~al.}(2011)\citenamefont
  {Romanzin}, \citenamefont {Ioppolo}, \citenamefont {Cuppen}, \citenamefont
  {van Dishoeck},\ and\ \citenamefont {Linnartz}}]{Ro11}%
  \BibitemOpen
  \bibfield  {author} {\bibinfo {author} {\bibfnamefont {C.}~\bibnamefont
  {Romanzin}}, \bibinfo {author} {\bibfnamefont {S.}~\bibnamefont {Ioppolo}},
  \bibinfo {author} {\bibfnamefont {H.~M.}\ \bibnamefont {Cuppen}}, \bibinfo
  {author} {\bibfnamefont {E.~F.}\ \bibnamefont {van Dishoeck}}, \ and\
  \bibinfo {author} {\bibfnamefont {H.}~\bibnamefont {Linnartz}},\ }\href@noop
  {} {\bibfield  {journal} {\bibinfo  {journal} {JCP}\ }\textbf {\bibinfo
  {volume} {134}},\ \bibinfo {pages} {084504} (\bibinfo {year}
  {2011})}\BibitemShut {NoStop}%
\bibitem [{\citenamefont {Smoluchowski}(1983)}]{Sm83}%
  \BibitemOpen
  \bibfield  {author} {\bibinfo {author} {\bibfnamefont {R.}~\bibnamefont
  {Smoluchowski}},\ }\href@noop {} {\bibfield  {journal} {\bibinfo  {journal}
  {JPC}\ }\textbf {\bibinfo {volume} {87}},\ \bibinfo {pages} {4229} (\bibinfo
  {year} {1983})}\BibitemShut {NoStop}%
\bibitem [{\citenamefont {Tielens}\ and\ \citenamefont
  {Allamandola}(2011)}]{Ti11}%
  \BibitemOpen
  \bibfield  {author} {\bibinfo {author} {\bibfnamefont {A.~G. G.~M.}\
  \bibnamefont {Tielens}}\ and\ \bibinfo {author} {\bibfnamefont {L.~J.}\
  \bibnamefont {Allamandola}},\ }in\ \href@noop {} {\emph {\bibinfo {booktitle}
  {Cool Interstellar Physics and Chemistry.}}}\ (\bibinfo  {publisher} {Pan
  Stanford Publishing Pte Ltd: Singapore},\ \bibinfo {year} {2011})\BibitemShut
  {NoStop}%
\bibitem [{\citenamefont {Tielens}\ and\ \citenamefont {Hagen}(1982)}]{Ti82}%
  \BibitemOpen
  \bibfield  {author} {\bibinfo {author} {\bibfnamefont {A.~G. G.~M.}\
  \bibnamefont {Tielens}}\ and\ \bibinfo {author} {\bibfnamefont
  {W.}~\bibnamefont {Hagen}},\ }\href@noop {} {\bibfield  {journal} {\bibinfo
  {journal} {A\&A}\ }\textbf {\bibinfo {volume} {114}},\ \bibinfo {pages} {245}
  (\bibinfo {year} {1982})}\BibitemShut {NoStop}%
\bibitem [{\citenamefont {Tomellini}(2013)}]{To13}%
  \BibitemOpen
  \bibfield  {author} {\bibinfo {author} {\bibfnamefont {M.}~\bibnamefont
  {Tomellini}},\ }\href@noop {} {\bibfield  {journal} {\bibinfo  {journal}
  {Physica A}\ }\textbf {\bibinfo {volume} {392}},\ \bibinfo {pages} {875T}
  (\bibinfo {year} {2013})}\BibitemShut {NoStop}%
\bibitem [{\citenamefont {Ward}\ and\ \citenamefont {Price}(2011)}]{Wa11}%
  \BibitemOpen
  \bibfield  {author} {\bibinfo {author} {\bibfnamefont {M.~D.}\ \bibnamefont
  {Ward}}\ and\ \bibinfo {author} {\bibfnamefont {S.~D.}\ \bibnamefont
  {Price}},\ }\href@noop {} {\bibfield  {journal} {\bibinfo  {journal} {ApJ}\
  }\textbf {\bibinfo {volume} {741}},\ \bibinfo {pages} {121} (\bibinfo {year}
  {2011})}\BibitemShut {NoStop}%
\bibitem [{\citenamefont {Watanabe}\ \emph {et~al.}(2010)\citenamefont
  {Watanabe}, \citenamefont {Kimura}, \citenamefont {Kouchi}, \citenamefont
  {Chigai}, \citenamefont {Hama},\ and\ \citenamefont {Pirronello}}]{Wa10}%
  \BibitemOpen
  \bibfield  {author} {\bibinfo {author} {\bibfnamefont {N.}~\bibnamefont
  {Watanabe}}, \bibinfo {author} {\bibfnamefont {Y.}~\bibnamefont {Kimura}},
  \bibinfo {author} {\bibfnamefont {A.}~\bibnamefont {Kouchi}}, \bibinfo
  {author} {\bibfnamefont {T.}~\bibnamefont {Chigai}}, \bibinfo {author}
  {\bibfnamefont {T.}~\bibnamefont {Hama}}, \ and\ \bibinfo {author}
  {\bibfnamefont {V.}~\bibnamefont {Pirronello}},\ }\href@noop {} {\bibfield
  {journal} {\bibinfo  {journal} {ApJL}\ }\textbf {\bibinfo {volume} {714}},\
  \bibinfo {pages} {L233} (\bibinfo {year} {2010})}\BibitemShut {NoStop}%
\bibitem [{\citenamefont {Wolff}\ \emph {et~al.}(2010)\citenamefont {Wolff},
  \citenamefont {Lohmar}, \citenamefont {Krug}, \citenamefont {Frank},\ and\
  \citenamefont {O.}}]{Wo10}%
  \BibitemOpen
  \bibfield  {author} {\bibinfo {author} {\bibfnamefont {A.}~\bibnamefont
  {Wolff}}, \bibinfo {author} {\bibfnamefont {I.}~\bibnamefont {Lohmar}},
  \bibinfo {author} {\bibfnamefont {J.}~\bibnamefont {Krug}}, \bibinfo {author}
  {\bibfnamefont {Y.}~\bibnamefont {Frank}}, \ and\ \bibinfo {author}
  {\bibfnamefont {B.}~\bibnamefont {O.}},\ }\href@noop {} {\bibfield  {journal}
  {\bibinfo  {journal} {Phys. Rev. E}\ }\textbf {\bibinfo {volume} {81}},\
  \bibinfo {pages} {061109} (\bibinfo {year} {2010})}\BibitemShut {NoStop}%
\end{thebibliography}%

\end{document}